\documentclass[prl,aps,floatfix,showpacs,tightenlines,twocolumn,superscriptaddress,amsmath,amssymb]{revtex4}

\usepackage{graphicx}
\usepackage{dcolumn}
\usepackage{bm}
\usepackage{epsfig}
\def\comment#1{}

\begin{document}

\title{From quantum fusiliers to high-performance networks}

\author{W. J. Munro}\email{bill.munro@hp.com}
\affiliation{National Institute of Informatics, 2-1-2 Hitotsubashi, Chiyoda-ku, Tokyo 101-8430, Japan}
\affiliation{Hewlett-Packard Laboratories, Filton Road, Stoke Gifford, Bristol BS34 8QZ, United Kingdom} 

\author{K. A. Harrison}
\affiliation{Hewlett-Packard Laboratories, Filton Road, Stoke Gifford, Bristol BS34 8QZ, United Kingdom} 

\author{A. M. Stephens}
\affiliation{National Institute of Informatics, 2-1-2 Hitotsubashi, Chiyoda-ku, Tokyo 101-8430, Japan}

\author{S. J. Devitt}
\affiliation{National Institute of Informatics, 2-1-2 Hitotsubashi, Chiyoda-ku, Tokyo 101-8430, Japan}

\author{Kae Nemoto}
\affiliation{National Institute of Informatics, 2-1-2 Hitotsubashi, Chiyoda-ku, Tokyo 101-8430, Japan}

\begin{abstract}
Our objective was to design a quantum repeater capable of achieving one million entangled pairs per second 
over a distance of 1000km. We failed, but not by much. In this letter we will describe the series of 
developments that permitted us to approach our goal. We will describe a mechanism that permits the creation 
of entanglement between two qubits, connected by fibre, with probability arbitrarily close to one and 
in constant time. This mechanism may be extended to ensure that the entanglement has high fidelity without 
compromising these properties. Finally, we describe how this may be used to construct a quantum repeater 
that is capable of creating a linear quantum network connecting two distant qubits with high fidelity. 
The creation rate is shown to be a function of the maximum distance between two adjacent quantum repeaters. 
\end{abstract}

\date{\today}

\maketitle

\section{Introduction}

The twentieth century saw the discovery of quantum mechanics,  a set of principles describing physical 
reality at the atomic level of matter. These principles have been used to develop much of today's advanced 
technology including, for example, today's microprocessors. Quantum physics also allows a new paradigm for 
the processing of information --- a field known as quantum information processing \cite{Nielsen00,spiller05}. 
Over the last decade we have seen a huge worldwide effort to develop and explore  
quantum-information based devices and technologies \cite{dowling02,spiller06}. Quantum key distribution (QKD) enabled 
devices are already commercially available \cite{gisin02}. The next step after this is likely to be small scale processors, 
probably distributed in nature. 

Quantum repeaters are a natural candidate to consider \cite{briegel98}. Their role 
is to enable the creation of entangled states between remote locations. Long-distance entanglement 
is achieved by placing a number of repeater nodes in-between two end points and creating entangled 
links between the adjacent nodes. Once a node has links both to the left and to the right, 
entanglement swapping within the nodes then allows longer-range entangled links to 
be formed. Once swapping operations have occurred at all the intermediate nodes an end-to-end entangled 
link will have been formed. These entangled pairs can then be used in QKD, quantum communication, or distributed 
quantum computation.

The current goal of many research groups is to produce a stream of entangled qubits over long distances, 
preferably with rates in the MHz range, There have been many proposals for how this could be achieved 
and a number of ''in-principle'' demonstrations have been performed. Such proposals have generally 
focused on the quantum components necessary to create entangled links between neighboring nodes, 
purification of these links, and swap operations to create longer-distance links \cite{sangouard10,loock06,munro08,childress06,enk98,duan01,chen07a}. 
The entangled links are generally created by entangling an optical signal (appendix 1) with a qubit and then 
transmitting that signal over a channel to the neighboring node. Here the signal entangles with a 
qubit within that node and then a measurement is made on the quantum signal indicating successful generation 
or not \cite{loock06,munro08}. The probability of successfully generating the link scales at best as 
$\exp [{-L/L_0}]$, where $L$ is the distance between repeater nodes and $L_0$ the attenuation length of the fiber. 

The next step is to look at the overall design of the repeater network, in terms of both the quantum 
and classical components. The communication time for classical messages to be transmitted between nodes 
severely limits the performance of a repeater network. Messages generally need to be sent  between nodes 
in all of the three key quantum stages of a repeater network: entanglement distribution, purification, 
and swapping. In this letter we will describe a pipe-lined architecture where one knows
when the end-to-end entangled pairs are going to be available. 

\section{Quantum Fusiliers and Fusilands}

The major issue affecting the performance for a quantum repeater is the probabilistic nature of the 
generation of entanglement between adjacent nodes and not knowing when such a link is going to be 
available. This issue means that a confirmation signal needs to be sent back 
from the receiver to the transmitter side and so the generation rate is ultimately
limited by this round trip transmission time. With typical repeater nodes being 
separated by, say, 40km this would take on the order of 400 $\mu$s. Now with the 
probability of success for entanglement generation being below 25\%, quite a number of attempts 
are going to be needed before we are ''{\it guaranteed}'' a link. A significant time delay results 
if the attempts are performed sequentially. One could parallelize the operations but with 
significantly more resources. One must be able to do better!

\begin{figure}
\begin{center}
\includegraphics[scale=0.4]{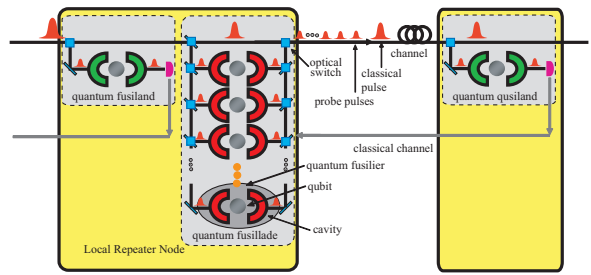}
\end{center}
\caption{Schematic representation of a quantum repeater node and its link to its next neighbor. 
The basic repeater is composed of two fundamental components: a quantum fusillade containing 
multiple fusiliers (transmission cavities each with a qubit within them) and a quantum fusiland 
(receiving cavity with a qubit within it and a signal detector).} 
\label{fig1}
\end{figure}

A simpler design does indeed exist which we depict in Fig.~(\ref{fig1}). 
In this design each repeater node comprises two fundamental  parts: a quantum 
fusillade containing multiple fusiliers (transmitters) and quantum fusilands 
(receivers). There are generally more fusiliers than fusilands and for the
moment we will consider a single fusiland. The creation of a constant-time entanglement link 
begins by a classical pulse initiating all the fusiliers in that node to 
prepare individual quantum optical signals. These signals then interact and become 
entangled with the qubits in their respective fusilier cavities. The signals then 
propagate, temporally multiplexed together with the classical heralding pulse, along the fiber to the 
fusiland in the next repeater node. The classical pulse announces to the fusiland that a series 
of quantum signals are about to arrive and so the fusiland initializes the 
qubit into the appropriate state and then interacts with the first fusilier's 
quantum signal. The signal is then measured to determine whether a successful 
entanglement-creation operation has occurred. If not, the fusiland qubit is re-prepared 
for the arrival of the second fusilier's signal and the same interaction/measurement procedure is
performed. This continues until a success is reported. A successful 
result triggers two operations: first, it stops any further signals interacting with the fusiland; 
and secondly, it dispatches a classical message back to the fusiland informing 
it of which fusilier was successful. The time taken from firing the fusillade to receiving the 
classical message is essentially the round trip time between two adjacent repeaters and is a constant.
With enough fusiliers we can ''{\it guarantee}'' the entangled link exists. The failure probability is given by 
$p_f=\left(1-p\right)^n$, where $n$ is the number of fusiliers and $p$ is the success probability 
of a single fusilier/fusiland. With $p=0.25$, 16 fusiliers are needed for $p_f<0.01$.

With only one receiving fusiland we have to discard any further fusilier signals once a measurement 
has been successful. This is obviously a waste, but we could utilize a few extra fusilands. 
Once the first fusiland has been successfully entangled, we then route the remaining signals 
to the next fusiland. When that's successful, we then go on to the next and so on. If we have $n$ 
fusiliers and $m$ fusilands then the probability that all $m$ links have not been established is 
$p_f(m)=\sum_{j=1}^{m} {n \choose {j-1}} p^{j-1} \left(1-p\right)^{n-j+1}$. For $p_f=0.01$ and $p=0.25$, 
the numbers of fusiliers/fusilands needed are (n=16,m=1), (n=24,m=2), (n=70,m=10), (n=485,m=100) 
which in the asymptotic limit of large $m$ goes to (n=m/p, m). This clearly shows the advantage 
of having multiple fusilands in terms of resource efficiency.

With multiple entanglement links available between adjacent nodes, there are various possibilities 
for how these can be used. The simplest is just to use them in parallel to improve the overall 
network performance, however as our entangled links may not be perfect we need to be able to purify 
them. Normal purification protocols are problematic since they are probabilistic and require two-way 
communication to determine if we have succeeded or failed \cite{dur07,pan01}. Upon failure our entangled 
links are destroyed and we must start the link generation again. This is a major performance issue but 
it can be solved by using quantum error correction \cite{devitt09}. 

The particular error-correction code to be used will depend on the errors induced in the entanglement 
generation process and on the failure rate of the quantum gates at each node. If we assume perfect local 
gates and that the predominant channel error (excluding loss) is a bit-flip error, then our entangled 
link can be represented by 
\begin{eqnarray}
\rho [F]= \frac{F}{2} |gg+ee\rangle \langle gg+ee|+\frac{1-F}{2} | ge+eg\rangle \langle eg+ge| ,
\end{eqnarray}
where $F$ measures the fidelity (quality) of the entangled pairs one is trying to create. 
In this case, to create an entangled pair with fidelity $F'>F$ we make use of a three-qubit 
repetition code, which corrects a single bit-flip error as follows: Given three copies of 
$\rho (F)$ we perform non-destructive parity measurements on the first and second and then 
on the second and third fusiliers, recording the results $p_{12}$ and $p_{23}$. The second 
and third fusiliers are then measured out in the $X$ basis. On the fusiland side identical 
parity measurements are performed with results, say, $r_{12}$ and $r_{23}$ and then the 
second/third fusilands are measured out in the $X$ basis. The resulting entangled state 
is $\rho [F'= F^3+3 F^2(1-F)]$ up to a bit-flip correction determined by $p_{12}$, $p_{23}$, 
$r_{12}$, and $r_{23}$ and a phase-flip correction determined by the results of the four $X$ 
measurements. These corrections simply update the Pauli frame and need only be communicated 
to one end of the network. This means we do not need to wait and so the fusiliers and fusilands 
can be further processed. 

This simple protocol is quite effective at increasing the fidelity of the remaining pair 
relative to that of the initial pairs; for instance if we started with $F=0.95$ we would 
have $F'\geq 0.99$. Importantly, the non-determinism inherent in purification-based schemes 
is not present in this scheme, allowing for pipe-lining of the overall repeater network. To 
extend entanglement beyond neighboring nodes we perform swap operations (achieved by parity 
gates) between local fusiliers and fusilands when the local fusilier is entangled to the right 
and the fusiland to the left. This removes those local fusiliers and fusilands and creates a 
longer range link. After the swap operation the quality of the new link is likely to have 
degraded and so more error correction may be required. 

In the case of a general channel error and faulty local gates, to achieve fault tolerance 
whilst keeping with the spirit our design we can simply replace physical qubits with logical 
qubits encoded with a concatenated code such as the Bacon-Shor code \cite{Bacon1}. Then error 
correction is performed at the same time as entanglement swapping without any need for additional protocols \cite{jiang09}. 
This contrasts with other recent schemes based on planar codes and cluster states \cite{Perseguers1, Perseguers2, Fowler1}.
Since logical Bell pairs are required to perform error correction, one promising approach is to produce 
many logical Bell pairs at each node, rejecting pairs when errors are detected, so that a high-quality 
pair is always available when required \cite{Knill1}. We expect that this will yield a scheme which 
has a high threshold ($>1$\%) for channel \emph{and} gate errors whilst retaining the deterministic 
nature of the protocol. As with all error-correction schemes, the maximum error rate that is 
tolerable will depend ultimately on the number of entangled links we have available, the number 
of qubits at each node, and our target fidelity. However, we are confident that our method for 
establishing entanglement between repeater nodes gives us the flexibility to tailor error correction 
to communication tasks to ensure high fidelity entanglement with a practical amount of resources.

\section{A quasi-asynchronous design}

\begin{figure}
\begin{center}
\includegraphics[scale=0.4]{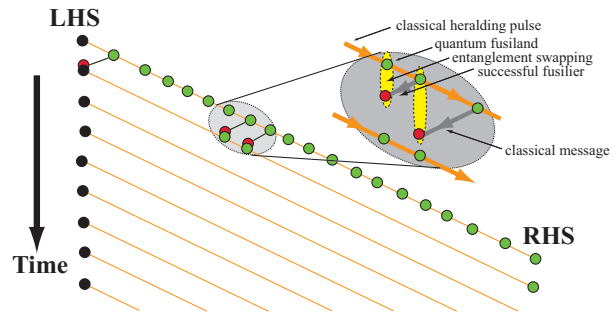}
\end{center}
\caption{Schematic representation of a quasi--asynchronous repeater  network. Entanglement generation is 
initiated on the left-hand side (LHS) where the system clock is located. In this design the classical heralding 
pulse from the left-hand most node propagates to the furthermost right-hand node (RHS) initiating all the 
fusiliers as it propagates. Swap operations within local nodes occur 
when the local fusiliers and fusilands have links to their neighbors. The left-hand node can start its next 
entanglement generation cycle after the round trip time for a entanglement generation between neighboring 
nodes. The classical heralding pulse on this next round picks up the Pauli frame information as it propagates 
through the network and makes it available to the right end node as it arrives.}
\label{fig2}
\end{figure}

With all the quantum components available we now need to consider appropriate strategies for 
putting this network together and how it will operate. This will need to involve both the quantum resources 
and the classical communication resources. The two logical choices for how such a network could operate are basically 
either a synchronous or asynchronous scheme. A synchronous scheme requires all the individual repeater nodes to 
have a shared clock which in certain circumstances could be challenging.  An asynchronous design does not 
require this and so it is the design we will focus on here.  We depict such a scheme in Fig.~(\ref{fig2}) and 
note that an advantage of this design is that the distance between adjacent nodes need not be the same 
(some could be at 10km say, others at 40km).

The quasi-asynchronous design begins with the clock in the left hand network node initiating the 
classical heralding pulse that is going to propagate along the whole network from left to right. As it goes it 
will initiate the fusiliers to fire the signals to the fusiland in the adjacent node and thus we will see the fusiliers 
firing in temporal progression from the left hand side of the network to the right hand side. Each of the adjacent nodes 
reports by a classical message which fusilands were successful and when that node has a link both to the left and 
the right the entanglement swap operation is performed, creating a longer distance link and freeing the fusiliers and 
fusilands in that node for future operations. The results of the parity measurements and swap operations are then 
available at that local node. We propagate this information on the heralding pulse for the next round of long-range entanglement generation. It is important that the next heralding pulse arrives at the repeater nodes after the 
swapping operations have been performed as the herald will pick this information up. It also means we know exactly 
when the entanglement link is ready to use and so we have an efficient pipe-lined design.

\section{A butterfly design}

As the entanglement generation is effectively flowing from left to right, the left-hand fusilier and the right-hand fusiland become entangled at quite different 
times. For QKD-like applications this is not an issue. For computational applications this could be an issue, but a simple solution is to split the network into two halves. The actual location of the 
split depends on the topology of the network, but is chosen to maximize throughput and to balance the 
availability of the left and right qubits. Each side is going to see a generalized 
parity for its half of the network. The two halves can be simply connected by entanglement swapping and this information 
propagated to either the left- or right-hand end with the next heralding pulse. It also means these resources in the 
''central'' node are freed relatively quickly and consequently we do not need exceptionally long lived qubits anywhere in 
the repeater network. This should significantly lessen the technological challenge inherent in distributed quantum 
information processing as the quantum resources now have to be good on time scales associated with the round 
trip time between adjacent nodes and not the propagation time over the whole network.

\section{Discussion}

We have so far presented a highly optimized design for a quantum repeater and its associated use in 
a network where the key element is the construction of a constant-time, near-deterministic, high fidelity 
entanglement link generator between neighboring repeater nodes. This time is of the order of the round 
trip time between adjacent nodes, approximately 0.4ms for a 40km link (0.1ms for a 10km link) and so allows 
a maximum rate/fusiland of 2500 (10000) entangled pairs between adjacent nodes. With more fusilands per 
repeater node one can approach a MHz rate. By utilizing oen way error correction the near-deterministic 
nature can be maintained without any significant time cost. Finally by utilizing a butterfly network 
design the end nodes becomes entangled at roughly the same time with the classical 
generalized parity results arriving one cycle (round trip) later. This allows a highly efficient and 
pipe-lined architecture. While we have considered a linear design the network topology can be easily 
generalized.. 

\noindent {\em Acknowledgments}: We would like to thank Clare Horsman and Tim Spiller for valuable  
discussions. This work was supported in part by MEXT and NICT in Japan and the EU project HIP.

\section{Appendix - Entanglement links}

One of the core elements necessary in any repeater design is the creation of entanglement 
between nearest neighbor links. This entanglement will be created between two electronic
spins placed in cavities at neighboring repeater stations with nuclear spins available for 
quantum storage. The electronic and nuclear-spin systems may be achieved, for example, 
by single electrons trapped in quantum dots, by neutral donor impurities in semiconductors 
or NV diamond centers. For a sufficient interaction between the electron and the light field, 
the system should be placed in a cavity resonant with the light. The mechanism for the 
entanglement between nodes generally fall into two categories.
\begin{itemize}
\item The heralded creation of very high fidelity entangled links utilising single photon or 
weak coherent sources generally with a low probability of success. The qubit-light 
field can operate in a number of regimes including on-resonance and dispersive. Moderate to strong 
coupling regimes are generally required. 
\item The creation of moderate fidelity entangled links ulitizing strong coherent fields and 
homodyne detection generally with a moderate to high probability of success. The qubit-light 
field generally operate in the dispersive regime.
\end{itemize}
Which is better to use really depends on the physical system but the second approach can use 
the same qubit-photon interacts for the local gate operations necessary in purification and 
entanglement swapping.


\begin{thebibliography}{99}

%
\bibitem{Nielsen00}
M. A. Nielsen and I. L Chuang, Quantum Computation and Quantum Information. Cambridge; New York: Cambridge University Press (2000).
%
\bibitem{spiller05} 
T. P. Spiller, W. J. Munro, S. D. Barrett, and P.Kok, Contemporary Physics {\bf 46}, 407 (2005).
%
\bibitem{dowling02}
J. P. Dowling and G. J. Milburn, Quantum Technology: The Second Quantum Revolution, arXiv:quant-ph/0206091 (2000).
%
\bibitem{spiller06}
T P Spiller and W J Munro, J. Phys.: Condens. Matter {\bf 18}, 1 (2006).
%
\bibitem{gisin02}  
N. Gisin, G. Ribordy, W. Tittel, and H. Zbinden, Rev. Mod. Phys. {\bf 74}, 145 (2002).
%
\bibitem{briegel98} 
H.-J. Briegel, W. D\"ur, J. I. Cirac, and P. Zoller,  Phys. Rev. Lett. {\bf 81}, 5932 (1998).
%
\bibitem{sangouard10} N. Sangouard, C. Simon, H. de Riedmatten, and N. Gisin, Quantum repeaters based on atomic ensembles and linear optics, arXiv:0906.2699v2 (2009) and references within.
%
\bibitem{loock06} 
P. van Loock, T. D. Ladd, K. Sanaka, F. Yamaguchi, Kae Nemoto, W. J. Munro, and Y. Yamamoto,  Phys. Rev. Lett. {\bf 96}, 240501 (2006).
%
\bibitem{munro08}W. J. Munro, R. Van Meter, Sebastien G. R. Louis, and Kae Nemoto, Phys. Rev. Lett. {\bf 101}, 040502 (2008).
%
\bibitem{childress06} 
L. Childress, J. M. Taylor, A. S. Sorensen, and M. D. Lukin, Phys. Rev. Lett. {\bf 96}, 070504 (2006).
%
\bibitem{enk98} 
S. J. Enk, J. I. Cirac, and P. Zoller, Science {\bf 279}, 205 (1998).
%
\bibitem{duan01} 
L.-M. Duan, M. D. Lukin, J. I. Cirac, and P. Zoller, Nature {\bf 414}, 413 (2001).
%
\bibitem{chen07a} 
B. Zhao, Z.-B. Chen, Y.-A. Chen, J. Schmiedmayer, and J.-W. Pan, Phys. Rev. Lett. 98, 240502 (2007).
%
\bibitem{dur07} W. D\"ur and H. J. Briegel, Rep. Prog. Phys. {\bf 70}, 1381 (2007).
%
\bibitem{pan01} 
J. Pan, C. Simon, C. Brukner, and A. Zeilinger, Nature {\bf 410}, 1067 (2001).
%
\bibitem{devitt09} S. J. Devitt, K. Nemoto, and W. J. Munro, The idiots guide to Quantum Error Correction, arXiv:0905.2794 (2009).
%
\bibitem{Bacon1} D. Bacon, Phys. Rev. A {\bf 73}, 012340 (2006).
%
\bibitem{jiang09} L. Jiang, J. M. Taylor, Kae Nemoto, W. J. Munro, R. Van Meter, and M. D. Lukin, Phys. Rev A {\bf 79}, 032325 (2009).
%
\bibitem{Perseguers1} S. Perseguers, L. Jiang, N. Schuch, F. Verstraete, M.D. Lukin, J.I. Cirac, and K.G.H. Vollbrecht, Phys. Rev. A {\bf 78}, 062324 (2008).
%
\bibitem{Perseguers2} S. Perseguers, Fidelity threshold for long-range entanglement in quantum networks, arXiv:0910.1459v1 (2009).
%
\bibitem{Fowler1} A. G. Fowler, D. S. Wang, T. D. Ladd, R. Van Meter, and  L. C. L. Hollenberg, Fast, fault-tolerant quantum communication, arXiv:0910.XXXXv1 (2009).
%
\bibitem{Knill1} E. Knill, Nature {\bf 434}, 39 (2005).
\end{thebibliography}
\end{document}